
\magnification\magstep1
\parskip=\medskipamount
\hsize=6 truein
\vsize=8.2 truein
\hoffset=.2 truein
\voffset=0.4truein
\baselineskip=14pt
\tolerance=500


\def\reals{\hbox{\bf R}}

\def\B{{\cal B}}
\def\H{{\cal H}}
\def\scri{\hbox{${\cal I}^+$}}

\def\Box#1{\mathop{\mkern0.5\thinmuskip
\vbox{\hrule
\hbox{\vrule
\hskip#1
\vrule height#1 width 0pt\vrule}%
\hrule}%
\mkern0.5\thinmuskip}}
\def\endproof{\Box{5pt}}

\font\titlefont=cmbx12
 at 10 truept
\font\authorfont=cmcsc10
\font\addressfont=cmsl10 at 10 truept
\font\smallbf=cmbx10 at 10 truept

\outer\def\beginsection#1\par{\vskip0pt plus.2\vsize\penalty-150
\vskip0pt plus-.2\vsize\vskip1.2truecm\vskip\parskip
\message{#1}\leftline{\bf#1}\nobreak\smallskip\noindent}

\newdimen\itemindent \itemindent=13pt
\def\textindent#1{\parindent=\itemindent\let\par=\resetpar%
\indent\llap{#1\enspace}\ignorespaces}

\let\oldpar=\par
\def\resetpar{\oldpar\parindent=0pt\let\par=\oldpar}

\font\ninerm=cmr9 \font\ninesy=cmsy9
\font\eightrm=cmr8 \font\sixrm=cmr6
\font\eighti=cmmi8 \font\sixi=cmmi6
\font\eightsy=cmsy8 \font\sixsy=cmsy6
\font\eightbf=cmbx8 \font\sixbf=cmbx6
\font\eightit=cmti8
\def\eightpoint{\def\rm{\fam0\eightrm}
  \textfont0=\eightrm \scriptfont0=\sixrm \scriptscriptfont0=\fiverm
  \textfont1=\eighti  \scriptfont1=\sixi  \scriptscriptfont1=\fivei
  \textfont2=\eightsy \scriptfont2=\sixsy \scriptscriptfont2=\fivesy
  \textfont3=\tenex   \scriptfont3=\tenex \scriptscriptfont3=\tenex
  \textfont\itfam=\eightit  \def\it{\fam\itfam\eightit}%
  \textfont\bffam=\eightbf  \scriptfont\bffam=\sixbf
  \scriptscriptfont\bffam=\fivebf  \def\bf{\fam\bffam\eightbf}%
  \normalbaselineskip=9pt
  \setbox\strutbox=\hbox{\vrule height7pt depth2pt width0pt}%
  \let\big=\eightbig \normalbaselines\rm}
\catcode`@=11 %
\def\eightbig#1{{\hbox{$\textfont0=\ninerm\textfont2=\ninesy
  \left#1\vbox to6.5pt{}\right.\n@space$}}}
\def\vfootnote#1{\insert\footins\bgroup\eightpoint
  \interlinepenalty=\interfootnotelinepenalty
  \splittopskip=\ht\strutbox %
  \splitmaxdepth=\dp\strutbox %
  \leftskip=0pt \rightskip=0pt \spaceskip=0pt \xspaceskip=0pt
  \textindent{#1}\footstrut\futurelet\next\fo@t}
\catcode`@=12 %


\bigskip
{\baselineskip=24 truept
\titlefont
\centerline{Is There a General Area Theorem for Black Holes?}
}

\vskip 1.1 truecm plus .3 truecm minus .2 truecm

\centerline{\authorfont Domenico Giulini}
\vskip 2 truemm
{\baselineskip=12truept
\addressfont
\centerline{Institut f\"ur theoretische Physik, Universit\"at Z\"urich}
\centerline{Winterthurerstrasse 190, CH-8057 Z\"urich, Switzerland}
\centerline{e-mail: giulini@physik.unizh.ch}
}
\vskip 1.5 truecm plus .3 truecm minus .2 truecm

\centerline{\smallbf Abstract}
\vskip 1 truemm
{\baselineskip=12truept
\leftskip=3truepc
\rightskip=3truepc
\parindent=0pt

{\eightpoint
The general validity of the area law for black holes is still  
an open problem. We first show in detail how to complete the usually 
incompletely stated text-book proofs under the assumption of piecewise 
$C^2$-smoothness for the surface of the black hole. Then we prove 
that a black hole surface necessarily contains points where it is 
not $C^1$ (called ``cusps'') at any time before caustics of the horizon 
generators show up, like e.g. in merging processes. 
This implies that caustics never disappear in the past and that  
black holes without initial cusps will never develop such.
Hence black holes which will undergo any non-trivial processes 
anywhere in the future will always show cusps. Although this does not 
yet imply a strict incompatibility with piecewise $C^2$ structures, it 
indicates that the latter are likely to be physically unnatural.
We conclude by calling for a purely measure theoretic proof of the 
area theorem.

\par}}

\beginsection{Introduction}

It seems to be widely accepted as fact that the surface area of a black
hole cannot decrease with time. However, the proofs offered in standard
text-books, like [HE], [MTW] and [W], are basically content with the remark
that this law follows from the non-convergence of the generators of the
future event horizon. It would indeed follow from this remark and
some elementary differential geometric considerations if the horizon
were a sufficiently smooth submanifold. Mathematically there is 
absolutely no reason why this should be true in general~[CG], which means 
that extra assumptions must be invoked (implicitly) in the text-books 
arguments. However, not much precise information seems to exist in the
literature concerning these extra assumptions. Perhaps the clearest 
statement is given in [CG], where the authors mention that 
in the text-book proofs of the area theorem ``something close to 
$C^2$ differentiability `almost everywhere' of the event horizon 
seems to have been assumed''. The text-books themselves do to mention 
any such condition. Below we show explicitly how to complete the 
text-book argumentation under the assumption of piecewise 
$C^2$-smoothness. But this clearly does not imply its necessity.

General considerations only prove the horizon to be locally 
Lipschitz continuous (denoted by $C^{1-}$)~[HE]. Mathematically this 
implies (pointwise) differentiability almost everywhere (with respect 
to the Lebesgue measures defined by the charts, see [F] Theorem~3.1.6), 
but it still allows the points of non-differentiability to be densely 
distributed~[CG]. Hence horizons exist which are nowhere $C^1$.
Given these mathematical facts, it is of interest to learn what 
{\it physical} conditions imply a breakdown of $C^1$ differentiability.
We prove that a black hole whose surface is $C^1$ at one time can 
never merge with other black holes and, more generally, never 
encounter new null generators for its future event horizon. In other 
words, an initial $C^1$-condition basically rules out any interesting 
physical process to happen in the future. Hence the only dynamically 
interesting holes for which the area law is actually proven are those 
whose surfaces are piecewise $C^2$ but not $C^1$. Presumably this 
class does not contain many, if any, physically realistic 
members. Cusps on the surfaces of colliding black holes can be clearly 
seen in numerical studies [L-W], but no analytic proof of their general 
existence seems to have been given so far.

Given the widely believed connection of the area law with 
thermodynamic properties of black holes on one side, 
and the widely expressed hope that this connection may be of 
heuristic value in understanding certain aspects of quantum gravity 
on the other, it seems important to know the most general conditions 
under which the area law is valid.

\beginsection{Notation, Facts and Assumptions} 

We assume 
space-time $(M,g)$ to be strongly asymptotically predictable 
(in the sense of [W]) and globally hyperbolic.
(It would be sufficient to restrict to a globally hyperbolic 
portion, as in Thm. 12.2.6 of [W].) $\scri$ (scri-plus)
denotes future null infinity, $J^-(\scri)$ its causal past and 
$B:=M-J^-(\scri)$ the black-hole region. Its boundary, 
$\partial B=:H$, is the future-event-horizon. $H$ is a closed, 
imbedded, achronal three-dimensional $C^{1-}$-submanifold of $M$ 
(Proposition 6.3.1 in [HE]). $H$ is generated by null geodesics
without future-endpoints. Past-endpoints occur only where 
null geodesics -- necessarily coming from $J^-(\scri)$ -- join onto 
$H$. Such points are called ``caustics''. Only at a caustic can a 
point of the horizon be intersected by more than one generator, and 
{\it all} generators that intersect a caustic enter the horizon at 
this point. Once a null geodesic has joined onto $H$ it will never 
encounter a caustic again (i.e. not intersect another generator) and
never leave $H$. See Box 34.1 in
[MTW] for a lucid discussion and partial proofs of these statements. 
Hence there are two different processes through which the area of a 
black hole may increase: First, new generators can join the horizon 
and, second, the already existing generators can mutually diverge.

Caustic points where $n$ (possibly infinite) new generators join 
in are said to be of multiplicity $n$. In [BK] it is proven that 
$H$ is not differentiable at $p$ iff $p$ is a caustic of multiplicity 
$n\geq 2$, and that caustics of multiplicity~1 are contained in 
the closure of those of higher multiplicity. In particular, $H$ 
cannot be of class $C^1$ at any caustic point. Points where $H$ 
is not $C^1$ will be called ``cusps''. By definition, being $C^1$ 
at $p$ implies that $H$ is differentiable 
in a whole neighbourhood $U$ of $p$. Conversely, it was shown that 
differentiability in some open neighbourhood $U$ of $p$ implies that 
$H$ is $C^1$ in $U$ ([BK], Prop. 3.3), so that the set of points where 
$H$ is $C^1$ is open. It follows that being $C^1$ at $p$ is in fact 
equivalent to being differentiable in some neighbourhood  of $p$. 
Hence the set of cusps is the closure of the set of points where 
$H$ is non-differentiable (caustics of multiplicity $\geq 2$) 
and hence also the closure of the set of all caustics.

Let $\Sigma$ be a suitably smooth (usually $C^2$) Cauchy surface, then 
$\B:=B\cap\Sigma$ is called the black-hole region at time $\Sigma$ 
and $\H:=H\cap\Sigma=\partial\B$ the (future-event-) horizon at time 
$\Sigma$. A connected component $\B_i$ of $\B$ is called a black-hole 
at time $\Sigma$. Its surface is $\H_i=\partial B_i$, which is a 
two-dimensional, imbedded $C^{1-}$-submanifold of $\Sigma$. 
We have seen that in general $\H$ may contain all kinds of 
singularities which would render standard differential geometric 
methods inapplicable. Adding the hypothesis of piecewise 
$C^2$-smoothness circumvents this problem.

By $exp:\, TM\rightarrow M$ we denote the exponential map. Recall that 
$exp_p(v):=\gamma(1)$, where $\gamma$ is the unique geodesic with 
initial conditions $\gamma(0)=p\in M$ and $\dot\gamma(0)=v\in T_p(M)$. 
For each $p$ it is well defined for $v$ in some open neighbourhood of 
$0\in T_p(M)$. One has $\gamma(t)=exp_p(tv)$. We shall assume the 
Lorentzian metric $g$ of $M$ to be $C^2$, hence the connection 
(i.e. the Christoffel Symbols) is $C^1$ and therefore the map 
$exp$ is also $C^1$. The last assertion is e.g. proven in~[L].

\beginsection{Local Formulation of the Area Law}

We consider two $C^2$ Cauchy surfaces with $\Sigma'$ to the future 
of $\Sigma$. The corresponding black-hole regions and surfaces 
are denoted as above, with a prime distinguishing those on $\Sigma'$.
We make the {\it assumption} that $\H$ is piecewise $C^2$, i.e.
each connected component $\H_i$ of $\H$ is the union of open subsets 
$\H^k_i$ which are $C^2$ submanifolds of $M$ and whose 2-dimensional 
measure exhaust that of $\H_i$: $\mu(\H_i-\bigcup_k\H_i^k)=0$, 
where $\mu$ is the measure on $\H$ induced from the metric $g$.

For each point $p\in\H^k_i$ there is a unique future- and 
outward-pointing null direction perpendicular to $\H_i^k$, which we 
generate by some future directed $l(p)\in T_p(M)$. We can choose 
a $C^1$-field $p\mapsto l(p)$ of such vectors over 
$\H_i^k$. The geodesics 
$\gamma_p:\,t\mapsto\gamma_p(t):=exp_p(tl(p))$ are generators of 
$H$ without future-endpoint. Therefore each $\gamma_p$ 
cuts $\Sigma'$ in a unique point $p'\in\H'$ at a unique parameter 
value $t=\tau(p)$. By appropriately choosing the affine 
parametrisations of $\gamma_p$ as $p$ varies over $\H_i^k$
we can arrange the map $\tau$ to be also $C^1$. Hence
$p\mapsto m(p):=\tau(p)l(p)$ is a null vector field of 
class $C^1$ over $\H_i^k$. We can now define the map 
$$
\Phi_i^k:\, \H_i^k\rightarrow\H'\,,\quad
         p\mapsto\Phi_i^k(p):=exp_p(m(p)),
\eqno{(1)}
$$
which satisfies the following  
\proclaim Lemma 1. $\Phi_i^k$ is (i) $C^1$, (ii) injective, 
(iii) non-measure-decreasing.

\noindent
(i) follows from the fact that the functions $m$ and $exp$ are $C^1$.
Injectivity must hold, since otherwise some of the generators of $H$ 
through $\H_i^k$ would cross in the future. By non-measure-decreasing
we mean the following: Let $\mu$ and $\mu'$ be the measures on $\H$ 
and $\H'$ induced by the space-time metric $g$. Then 
$\mu[U]\leq\mu'[\Phi_i^k(U)]$ for {\it each} measurable $U\subset\H_i^k$. 
Assuming the weak energy condition, this is a consequence of the 
{\it nowhere} negative divergence for the future geodesic congruence 
$p\mapsto\gamma_p$ (Lemma 9.2.2 in [HE]), as we will now show.

\noindent
{\sl Proof of (iii):} Set $H_i^k:=\bigcup_{p,t}exp_p(tl(p))$,
$\forall p\in\H_i^k$ and $\forall t\in R_+$, which is a 
$C^1$-submanifold of $M$ (the future of $\H_i^k$ in $H$). 
Let $l$ be the unique 
(up to a constant scale) future directed null geodesic 
(i.e. $\nabla_ll=0$) vector field on $H_i^k$ parallel to the generators.
Then $0\leq\nabla_{\mu}l^{\mu}=\pi^{\mu}_{\nu}\nabla_{\mu}l^{\nu}$,
where $\pi$ denotes the map given by the $g$-orthogonal projection 
$T(M)\vert_{H_i^k}\rightarrow T(H_i^k)$, followed by the quotient map 
$T(H_i^k)\rightarrow T(H_i^k)/\hbox{span}\{l\}$. Note that tangent 
spaces of $C^1$-cross-sections of $H_i^k$ at the point $p$ are 
naturally identified with $T_p(H_i^k)/\hbox{span}\{l(p)\}$. 
Since $\pi^{\mu}_{\nu}l^{\nu}=0$, we also have 
$\pi^{\mu}_{\nu}\nabla_{\mu}k^{\nu}\geq 0$ for 
$k=\lambda l$ and {\it any} $C^1$-function 
$\lambda:H_i^k\rightarrow\reals_+$. Hence this inequality is valid 
for any future pointing $C^1$-vector-field $k$ on $H_i^k$ parallel to the 
generators.  Given that, let $t\mapsto\phi_t$ be the flow
of $k$ and $A(t):=\mu_t[\phi_t(U)]:=\int_{\phi_t(U)}d\mu_t$, then
${\dot A}(t)=
\int_{\phi_t(U)}\pi^{\mu}_{\nu}(t)\nabla_{\mu}k^{\nu}(t)\,d\mu_t\geq 0$,
where $\pi(t)$ projects onto $T(\phi_t(\H_i^k))$,
$k(t)={{d}\over{dt'}}\vert_{t'=t}\phi_{t'}$ and $\mu_t=$ measure 
on $\phi_t(\H_i^k)$. Now choose $k$ such that $\phi_{t=1}=\Phi_i^k$,
then $\mu'[\Phi_i^k(U)]-\mu[U]=\int_0^1dt\dot{A}(t)\geq 0$.
\hfill $\endproof$

Part (iii) of Lemma~1 is the local version of the area law. 
By turning it into a global statement about areas one usually abandons 
some of its information. The most trivial global implication is
that the total sum of areas cannot decrease. A more refined version 
is as follows: Recall that black holes cannot bifurcate in the 
future (Proposition 9.2.5 of~[HE]). Hence all surface elements 
$\H_i^k$ of the $i$-th black-hole at time $\Sigma$ are mapped 
via $\Phi_i^k$ into the surface ${\H'}_i$ of a {\it single} black-hole at 
time $\Sigma'$. We call ${\H'}_i$ (i.e. the connected component 
of $\Sigma'\cap H$ into which $\H_i$ is mapped) the development 
of $\H_i$ at time $\Sigma'$. Lemma~1 now implies that its area 
cannot be less than that of $\H_1$. The non-bifurcation result 
implies that if the number $N'$ of black holes at time $\Sigma'$
is bigger than the number $N$ at time $\Sigma$, then there is 
an intermediate formation of $K\geq N'-N$ new black holes.
That these black-holes are `new', i.e. not present at time $\Sigma$, 
means that all generators of $H$ which intersect 
${\H'}_1\cup\cdots\cup{\H'}_K$ must have past-endpoints somewhere 
between $\Sigma$ and $\Sigma'$. A black hole at time $\Sigma'$ 
which is smaller than any black hole at time $\Sigma$ must also be 
new in this sense. Hence one way to express an area law would be 
as follows:

\proclaim Assertion (Area Law).  Consider two Cauchy surfaces, 
$\Sigma$ and $\Sigma'$, with $\Sigma'$ to the future of $\Sigma$.
Then the area of the development ${\H'}_i$ of any ${\H}_i$ cannot be 
less than that of $\H_i$. In particular, black holes at time $\Sigma'$ 
whose area is smaller than that of any black hole at time $\Sigma$
must have been formed in the meantime.

\noindent
Presently we do not have a proof that this statement is true in general.
But since it is a statement about measures, we suggest that it should 
be possible to give a proof without invoking fiducial (and probably 
irrelevant) differentiability assumptions.

\beginsection{Consequences}

The foregoing discussion allows to show that black holes whose 
surface is $C^1$ at one instant cannot undergo any non-trivial 
change in the future, like merging  processes or any other process
involving the incorporation of new generators. This we shall now 
explain in more detail. Let $\H_1$ be the surface a black hole at time 
$\Sigma$. We assume $\H_1$ to be a compact, 2-dimensional $C^1$-submanifold.
As before, $\Sigma'$ is to the future of $\Sigma$ and $\H'=H\cap\Sigma'$. 
We can construct a map $\Phi_1:\H_1\rightarrow\H'$, just analogous to the 
construction of $\Phi_i^k$ above, but now defined on {\it all} of $\H_1$. 
The $C^1$-condition on $\H_1$ now implies that $\Phi_1$ is $C^0$. 
$\Phi_1$ is also injective for the same reason as given for 
$\Phi_i^k$. Since $\H_1$ is connected, its image under $\Phi_1$ 
is also connected. Let ${\H'}_1\subset\H'$ be the connected part 
containing the image of $\Phi_1$. We show 

\proclaim Lemma 2. $\Phi_1:\,\H_1\rightarrow {\H'}_1$ is a 
homeomorphism.

\noindent
{\sl Proof.}
$\Phi_1$ is a closed map, because   
if $U\subset\H_1$ is closed $\Rightarrow$ $U$ is compact (since $\H_1$
is compact) $\Rightarrow$ $U':=\Phi_1(U)$ is compact (since $\Phi_1$
is continuous) $\Rightarrow$ $U'$ is closed (since ${\H'}_1$ is 
Hausdorff). From this follows that $\Phi_1$ is a homeomorphism 
onto its image. But $\Phi_1$ is also open. This follows directly from 
Brouwer's theorem on the invariance-of-domain, which states that 
any continuous injective map from an open $X\subset R^n$ into $R^n$ 
is open (Proposition 7.4 in [D]). This clearly generalizes to manifolds. 
Hence the image $\Phi_1(\H_1)\subset {\H'}_1$ is open, closed and 
connected, and hence all of ${\H'}_1$.
\hfill $\endproof$

Surjectivity of $\Phi_1$ implies that all generators of $H$ which 
intersect ${\H'}_1$ also intersect $\H_1$. Hence nowhere in its future 
will $\H_1$ be joined by new generators. Similar to the definition of 
$H_i^k$ above, let $H_1\subset H$ denote the future of $\H_1$ in $H$; 
then it follows that $H_1$ is free of caustics and therefore $C^1$. 
Hence we have 

\proclaim Proposition 1. Let the surface of a black hole at time 
$\Sigma$ be without cusps. Then this black hole will never 
encounter cusps to the future of $\Sigma$, in particular, it will
not merge with other black holes. 

\noindent
Another equivalent formulation, emphasizing that cusps will not die 
out in the past, is as follows:

\proclaim Proposition 2. At no time to the past of a cusp on $H$ 
will the surface of a black hole be without cusps. 

It is sometimes suggested that caustics just exist for some finite 
time interval during which the actual processes take place, like 
collision and coalescence of black holes or the infall of matter 
through the horizon (see e.g. [MTW]~34.5). Proposition~2 shows that 
this is not quite the right picture.

\beginsection{References}

\item{[BK]}
J.K. Beem and A. Kr\'olak:
Cauchy horizon endpoints and differentiability.
\hfill\break
gr-qc/9709046

\item{[CG]}
P.T. Chru\'sciel and G.J. Galloway:
Horizons non-differentiable on a dense set.
{\it Comm. Math. Phys.} {\bf 193} (1998), 449-470.

\item{[D]}
A. Dold: {\it Lectures on algebraic topology},
Springer Verlag, Berlin, 1972.

\item{[F]}
H. Federer: {\it Geometric measure theory}, 
Springer Verlag, New York, 1969.

\item{[HE]}
S.W. Hawking and G.F.R. Ellis:
{\it The large scale structure of space-time},
Cambridge University Press, 1973.

\item{[L]}
S. Lang:
{\it Differential Manifolds},
Springer-Verlag, Berlin, 1985.

\item{[L-W]}
J. Libson, J. Mass\'o, E. Seidel, W.-M. Suen and P. Walker:
Event horizons in numerical relativity: Methods and tests.
{\it Phys. Rev. D} {\bf 53} (1996), 4335-4350.

\item{[MTW]}
C.W. Misner, K.S. Thorne and J.A. Wheeler: {\it Gravitation},
W.H. Freeman and Company, San Francisco, 1973.

\item{[W]}
R.M. Wald: {\it General Relativity},
The University of Chicago Press, Chicago and London, 1984.

\end